\documentclass[12pt,oneside]{article}

\usepackage{cite}
\usepackage{acronym}
\usepackage{enumerate}  
\usepackage{a4wide}
\usepackage[left=2.5cm, right=2.5cm, top=2.5cm, bottom=2.5cm]{geometry}
\usepackage{fancyhdr}
\usepackage{graphicx}
\usepackage {lmodern} 
\usepackage{blindtext}
\usepackage{multirow}
\usepackage{amsmath}
\usepackage{amssymb}
\usepackage{mathrsfs}
\usepackage{bbm}
\usepackage{ragged2e}

\usepackage[T1]{fontenc}
\usepackage[utf8]{inputenc}
\usepackage[bookmarks]{hyperref}
\usepackage[justification=centering]{caption}
\usepackage{csquotes}
\usepackage[ngerman]{cleveref}
\usepackage[makeroom]{cancel}

\usepackage{tikz}
\usetikzlibrary{calc}
\usetikzlibrary{decorations.pathmorphing}
\tikzset{snake it/.style={decorate, decoration=snake}}
\usepackage{pgfplots}
\pgfplotsset{compat=1.18}
\usepgfplotslibrary{colorbrewer}

\title{Local Expansion Mechanisms \\ for Quantum-Scale Wormholes}

\author{{Philipp Dorau}\footnote{\href{mailto:philipp.dorau@uni-leipzig.de}{philipp.dorau@uni-leipzig.de}}$\quad\; \textrm{\&}\quad\;${Albert Much}\footnote{\href{mailto:much@itp.uni-leipzig.de}{much@itp.uni-leipzig.de}}\\[10pt]
\normalsize\textit{Institut f\"ur Theoretische Physik, Universit\"at Leipzig, 04103 Leipzig, Germany}}

\date{May 27, 2026}

\begin{document}

\maketitle

\vspace{0.37cm}

\begin{abstract}
Quantum models of spacetime have been conjectured to hypothetically allow for the formation of Planck scale wormholes. Building on the proposal of Morris, Thorne, and Yurtsever that such microscopic spacetime structures might be enlarged to macroscopic size, we revisit Roman’s analysis of a wormhole in an inflationary de Sitter background. In this context, we introduce a refined quasi-local toy mechanism, which we call the \emph{local inflation bubble}. This construction inflates a compact region of spacetime and thereby magnifies the underlying microstructure. Using the Einstein equations we determine the required stress-energy to sustain the bubble and obtain intrinsic lower bounds for the corresponding energy density, while acknowledging the continued reliance on exotic matter.
\end{abstract}

\newpage
\section{Introduction}\label{Introduction}
The possibility of time travel within the framework of general relativity has long intrigued both physicists and science‑fiction writers. While some frameworks require superluminal velocities, most standard approaches rely on wormholes, i.e. topological shortcuts between distant spacetime regions, that hypothetically enable the construction of time machines, as explained in \cite{MTY:1988wh}. \\

More specifically, a relatively simple model of a static and spherically symmetric \emph{traversable} wormhole was originally introduced by Morris and Thorne \cite{MT:1988wh} as a geometric construction connecting two distinct asymptotically flat regions of spacetime through a throat of minimal areal radius. The wormhole geometry is conveniently   expressed in terms of the proper radial length coordinate $l$,  which directly measures the physical radial distance as determined by local observers. In terms of $l$, the metric takes the form \cite{MG:2009cat}
\begin{equation}
\label{WormholeMetric}
ds^2 = - dt^2 + dl^2 + \left(s_0^2 + l^2\right) d\Omega^2,
\end{equation}

\noindent so that the wormhole throat is located at $l = 0$, while the limits $l \to \infty$ and $l \to -\infty$ correspond to the two distinct asymptotically flat regions. This parametrization is particularly advantageous for a geometric analysis, as it describes the wormhole as a smooth, horizonless bridge connecting two spatial infinities, with the minimal surface of radius $s_0$ at $l = 0$. \\

Most prominently, in order to solve the Einstein equations, the wormhole geometry \eqref{WormholeMetric} requires exotic matter, i.e. stress-energy tensors that violate the standard energy conditions. More precisely, the corresponding Einstein tensor is diagonal, with the temporal and radial components coinciding at the everywhere negative value
\begin{equation}
\label{MorrisThorneEinsteinTensor}
G_{tt} = G_{ll} = -\frac{s_0^2}{\left( s_0^2 + l^2 \right)^2}.
\end{equation}

\noindent Interpreting $T_{tt}$ as the energy density and $T_{ll}$ as the proper radial pressure of the (exotic) matter distribution, $G_{tt} < 0$ implies an everywhere negative energy density throughout the spacetime, thereby violating the weak, null, and averaged null energy conditions \cite{MT:1988wh,MTY:1988wh,HV:1998dw,VKD:2003prl,VKD:2004wq,FR:2005wh,MV:2017ec,KS:2020ec}.  
Classically, these properties are generally ruled out for physically reasonable matter sources, see, e.g. \cite{MV:2017ec,KS:2020ec}. Beyond conventional general relativity, however, within extended theories of gravity, traversable wormholes have been shown to exist as stable solutions supported by effective geometric contributions rather than exotic matter, see, e.g. \cite{SB:2021mod,dFBCdL:2021wh,dFC:2023wh}. \\

When traversable wormholes were first considered as potential pathways to interstellar travel, Morris, Thorne, and Yurtsever proposed that a sufficiently advanced civilization might be able to exploit quantum properties of matter at microscopic scales to create wormholes and then enlarge them to a practicable size \cite{MTY:1988wh}, see also \cite{FR:2005wh}. Indeed, quantum descriptions of matter, such as quantum field theory (QFT), generally allow for both local and averaged violations of the energy conditions, see \cite{PMG:1986casimir,BMM:2001casimir,FR:1995qei,FS:2008ei, MV:2017ec,KS:2020ec}. These violations are made precise in terms of quantum energy inequalities (QEIs), which were first established by Ford and Roman \cite{FR:1995qei}, who provided a (state-independent) limit for the averaged negative energy density $\rho$ of a quantized, free, minimally coupled, massless scalar field in static spacetimes by \cite{FR:1995qei}
\begin{align}\label{eq1}
\rho \geq -\frac{3\hbar c^3}{32\pi^2 t_{0}^4},
\end{align}

\noindent where $t_0$ denotes the characteristic timescale over which the average is taken. By comparison, the negative energy density derived from equation \eqref{MorrisThorneEinsteinTensor}, which is required to sustain a Morris-Thorne wormhole is given by
\begin{align}\label{eq2}
\rho = -\frac{c^4}{8\pi G}\frac {s_{0}^2}{(s_{0 }^2+l^2)^2}.
\end{align}

\noindent Equating the QEI bound \eqref{eq1} with the wormhole energy density \eqref{eq2} straightforwardly yields a general lower limit on the throat size $s_0$. In particular, if we assume $t_0$ to be the Planck time $t_P$, we obtain the specific minimal value
\begin{align}
s_0^2 = \frac{4}{3} \ell_{P}^{2},
\end{align}

\noindent where $\ell_{P}$ denotes the Planck length. \\

However, there exists a further, more stringent, restriction that is particularly relevant for wormhole geometries in semiclassical gravity, namely the achronal averaged null energy condition (achronal ANEC). In its standard form, the ANEC requires that the integral of the (renormalized) stress-energy tensor along a complete null geodesic curve be non-negative \cite{GO:2007aanec,Kontou:2024whr}. While this condition is known to be violated by quantum fields in curved spacetimes \cite{Visser:1996anecv}, there is weaker but more physically motivated version has been proposed, which only applies to null geodesics that are both complete and achronal, see \cite{GO:2007aanec}. More precisely, the achronal ANEC asserts that there exists no semiclassical spacetime in which the Einstein equations are fulfilled with a renormalized stress-energy tensor whose null average is negative along a complete achronal null geodesic. In fact, no violations of this condition are currently known, and it has been shown to be a sufficient condition to rule out causality violations, including time machines built from traversable wormholes, within the regime of validity of semiclassical gravity \cite{GO:2007aanec,Kontou:2024whr}. Consequently, the achronal ANEC provides a powerful obstruction to the possibility of wormholes and time machines in self-consistent semiclassical spacetimes. \\

Nonetheless, another hypothetical origin for microscopic wormholes lies in the quantum structure of spacetime itself: In the \emph{quantum foam} picture, see \cite{Wheeler:1957qf,Carlip:2023qf}, Planck scale fluctuations render the geometry highly irregular rather than smooth. Within such a framework, significantly fewer restrictions are currently known regarding their potential applications in the construction of causality violating configurations. Accordingly, it has been conjectured that microscopic wormholes may spontaneously emerge from the foam\footnote{We also refer to \cite{Stojkovic:2020form} for a related discussion on microscopic wormhole formation.} and, given an appropriate amplification mechanism, hypothetically be enlarged to a macroscopic size \cite{Roman:1993wh,Visser:1996wh}. In this work, we adopt this heuristic paradigm of the quantum foam to motivate the possibility of wormhole formation at the Planck scale. \\

Moreover, spacetime geometry is expected to become noncommutative around the Planck scale, at least effectively \cite{FMB:2023nc}. For such noncommutative spacetime models, there exist analogous constraints to the QEI \eqref{eq1}, see \cite{GM:2025ncqei}. These constraints suggest that negative energy densities are similarly permitted only over finite time scales. Hence, beyond the challenge of enlarging a Planck scale wormhole to a macroscopic size, this poses the additional problem of maintenance, i.e. sustaining the wormhole over a controlled finite time interval. \\

Specific mechanisms that primarily address the enlargement of microscopic wormholes, including both natural and artificial scenarios, have already been explored in the literature. Regarding a natural possibility, Roman \cite{Roman:1993wh} suggested that Planck scale physics in the early universe could have produced primordial wormholes, with cosmic inflation subsequently serving as a natural mechanism to enlarge them to macroscopic size. It was then shown that, in this setting, wormholes embedded in a cosmological de Sitter background exhibit positive energy densities at sufficiently late times, and that pointwise violations of the weak energy condition (WEC) at the throat decrease with time \cite{Roman:1993wh}. Alternatively, false vacuum bubbles, see \cite{KSSM:1981fvb,BGG:1987fvb}, offer a potential laboratory mechanism by which said advanced civilization could artificially enlarge a pre-existing, quantum-produced wormhole. More specifically, a false vacuum bubble consists of a compact nonvacuum region (the false vacuum), bounded by a thin domain wall and embedded in an exterior true vacuum spacetime \cite{BGG:1987fvb}. To exterior observers the solution appears as a Schwarzschild black hole, whereas interior observers perceive a closed inflating universe \cite{BGG:1987fvb}. \\

In this work, we introduce a refined toy mechanism, the \emph{local inflation bubble}\footnote{The terminology is  adapted from the notion of false vacuum bubbles, see \cite{KSSM:1981fvb,BGG:1987fvb}.}, which combines the features of the aforementioned scenarios. More specifically, the local inflation bubble consists of a compact inflationary region that locally enlarges (and maintains) a microscopic wormhole, while ensuring smooth transitions to the flat exterior without any sharp boundary. In mathematical terms, the geometry of a pure local inflation bubble is given by a smooth and \emph{compactly supported} deformation of Minkowski spacetime, described by the metric
\begin{equation}\boxed{
\label{BubbleMetric}
ds^2 = -dt^2 + e^{2f(t,r)} \left( dr^2 + r^2 d\Omega^2 \right),}
\end{equation}

\noindent where $f\in C_0^\infty(\mathbb{R}\times \mathbb{R}_+)$ is a smooth function with compact support in both $t$ and $r$. The geometry thus creates finite operational windows during which a local structure, such as a microscopic wormhole, is enlarged, while all associated stress-energy components remain bounded by construction. Outside of the bubble, the metric coincides with the original background geometry of Minkowski spacetime, allowing for a smooth embedding into an asymptotically flat environment. Regarding maintenance, the rapid local inflation of a microscopic wormhole may offer a first step towards prolonging its lifetime, by temporarily lifting it to macroscopic size, in analogy to cosmological inflation, which amplifies quantum fluctuations into persistent large-scale structures. However, we regard this as a preliminary idea rather than a resolution of the maintenance problem, and a definitive analysis should be performed in a subsequent work, not here.  This construction serves as a controlled \emph{Gedankenexperiment} for probing the geometric constraints, energy requirements, and favorable configurations of smooth localized inflation of Planck scale wormholes. \\

In Section \ref{GeometricConstruction}, we begin with a detailed geometric analysis of the metric \eqref{BubbleMetric}, and compute the relevant associated curvature quantities. This is followed by a short discussion of its physical viability in terms of the standard energy conditions and the positivity of total energy measured by static observers on a given constant-time surface. Moreover, in Section \ref{SecEstimates} we propose a concrete model of a local inflation bubble. More precisely, we consider the metric \eqref{BubbleMetric} for which the function $f\in C_0^\infty(\mathbb{R}\times\mathbb{R}_+)$ is given by a symmetric bump function of temporal and radial widths $\Delta t$ and $\Delta r$, respectively. Employing two different physically motivated assumptions for the parameter regimes, we derive estimates for the relevant energy scales required to the create such a bubble. Setting aside, for our present purposes, the nontrivial challenge of manipulating spacetime curvature at will, we then use these estimates to quantify \emph{how} advanced a civilization would have to be in order to realize such a mechanism successfully. \\

At last, in Section \ref{WormholeSection}, we turn to the explicit analysis of a Morris-Thorne wormhole embedded within a local inflation bubble. Specifically, this scenario is described by the metric
\begin{align} \boxed{
\label{EmbeddedWormholeMetric}
ds^2 = -dt^2 + e^{2f(t,l)} \left( dl^2 + \big(s_0^2 + l^2 \big)  d\Omega^2 \right),}
\end{align}

\noindent which physically corresponds to a temporarily expanding wormhole throat, in the spirit of \cite{Roman:1993wh}. Here again, we analyze the associated curvature quantities and extract stress-energy which is required to sustain this geometry. Beyond the observation that all such energy densities are intrinsically bounded from below, we identify a novel possibility of rendering the energy density at the wormhole throat positive.

\section{Geometric Construction of Local Inflation Bubbles}\label{GeometricConstruction}
Before studying the local inflation of a microscopic wormhole, let us consider the spacetime geometry of a pure local inflation bubble, which is contained in the Lemaître-Tolman class of metrics, see \cite{Lemaitre:1933metric,Tolman:1934metric,Bondi:1947metric,GP:2009st}. In our specific case, the model describes an isotropic but spatially inhomogeneous expansion that is confined to a compact region in spacetime. In comoving coordinates corresponding to a static observer, the line element of interest takes the form  
\begin{equation}\boxed{
ds^2 = -dt^2 + e^{2f(t,r)} \left( dr^2 + r^2 d\Omega^2 \right)}
\label{BubbleMetric2}
\end{equation}

\noindent where $d\Omega^2 = d\vartheta^2 + \sin^2\vartheta\,d\varphi^2$ denotes the canonical metric line element of the unit sphere $\mathbb{S}_1^2$, and $f \in C_0^\infty(\mathbb{R}\times\mathbb{R}_+)$ is a smooth, compactly supported, spherically symmetric function. The support of $f$ determines the spacetime region undergoing expansion, since outside of $\mathrm{supp}(f)$, the geometry coincides with flat Minkowski spacetime. \\
 
The choice of a strictly Minkowski background in the construction \eqref{BubbleMetric2} allows us to isolate the geometric and energetic content of the compactly supported deformation itself, without intertwining it with the curvature or horizon structure of a non-trivial background geometry. In particular, we aim to compute quantities such as the ADM mass, the Brown-York quasi-local mass, and the Newman-Penrose scalar $\Psi_4$, which acquire their standard, unambiguous interpretations only in an asymptotically flat setting. Since $f \in C_0^\infty(\mathbb{R}\times\mathbb{R})$ has compact support, the bubble probes scales far below any cosmological curvature radius, so that the ambient curvature of a realistic background (e.g. de Sitter during inflation, or a slowly varying semiclassical geometry) is negligible on $\mathrm{supp}\,f$. In this sense, the Minkowski construction captures the local behaviour of an isolated bubble, independently of the large-scale embedding that motivated scenarios such as Roman's inflating wormholes \cite{Roman:1993wh} or quantum-foam enlargement \cite{Wheeler:1957qf}. \\

We now briefly investigate the most important geometric properties of the bubble metric \eqref{BubbleMetric2}\footnote{In appendix \ref{AppendixCurvatureResults}, we provide exact results for the most relevant curvature quantities associated to the metric \eqref{EmbeddedWormholeMetric}, which includes the metric \eqref{BubbleMetric2}.}. Most notably, the spacetime is exactly flat outside the support of the function $f$, from which it follows immediately that the spacetime is asymptotically flat. Furthermore, given that $f \in C_0^\infty(\mathbb{R}\times\mathbb{R}_+)$, it follows by the extreme value theorem that it is necessarily bounded, and hence, both $e^f$ and $e^{-f}$ are bounded as well. This implies that there exist positive constants $A, B > 0$ such that  
\begin{equation}
\label{ConditionAbove}
A \, e^{2f(0,r)} \leq e^{2 f(t,r)} \leq B\, e^{2f(0,r)}
\end{equation}

\noindent for all $(t,r)\in \mathbb{R}^2$. Thus, the assumptions of \cite[Theorem 4.2]{FMOe:2020gh} are satisfied, so that that the spacetime $(\mathbb{R}^4, g)$ is globally hyperbolic if and only if $(\Sigma_t, h)$ is a complete Riemannian manifold, where $\Sigma_t \cong \mathbb{R}^3$ denotes a hypersurface of constant $t$ with induced spatial metric $h=e^{2f}\delta_{ij}$. This is indeed the case due to the boundedness of $e^{2f}$ and $e^{-2f}$, which implies that the identity map $(\mathbb{R}^3, e^{2f} \delta_{ij}) \rightarrow (\mathbb{R}^3, \delta_{ij})$ is bi-Lipschitz, i.e. there exist constants $L_1, L_2 > 0$ such that  
\begin{equation}
L_1 \, \delta_{ij} \ \leq\ e^{2f} \delta_{ij} \ \leq\ L_2 \, \delta_{ij},
\end{equation}
see \cite{BBI:2001mg}, similar to condition \eqref{ConditionAbove} above. Since Euclidean space $(\mathbb{R}^3, \delta_{ij})$ is complete and completeness is preserved under bi-Lipschitz equivalence \cite{BBI:2001mg}, it follows that $(\mathbb{R}^3, e^{2f} \delta_{ij})$ is complete, as well. Consequently, the spacetime $(\mathbb{R}^4, g)$ is globally hyperbolic. \\

To characterize the gravitational energy content of the spacetime, we consider two standard notions of quasi-local mass in spherically symmetric spacetimes, namely the Misner-Sharp-Hernandez (MSH) mass, and the Brown-York (BY) mass. Given the general definition of the MSH mass \cite{MS:1964mass,HM:1966mass}
\begin{equation}
M_\mathrm{MSH} := \frac{R}{2}\left( 1 - \nabla^a R \, \nabla_a R \right)
\end{equation}

\noindent for arbitrary spherically symmetric spacetimes with areal radius $R$, we obtain \cite{Mathematica}
\begin{equation}
M_\mathrm{MSH} = \frac{r^2}{2} e^f \left( r e^{2f} (\partial_t f)^2 - 2 \, (\partial_r f) - r (\partial_r f)^2 \right).
\end{equation}

\noindent The BY mass, on the other hand, is defined by the surface integral over a sphere of constant radius $r$ \cite{Poisson:2004tool}
\begin{equation}
\label{BrownYorkMass}
M_\textrm{BY} := \frac{1}{8\pi} \int_{\mathbb{S}^2_r} \left( K_0 - K_g \right) \,d\mathrm{vol}_{\mathbb{S}^2_r},
\end{equation}

\noindent where $K_g$ denotes the trace of the extrinsic curvature of the sphere $\mathbb{S}^2_r$ as embedded in the locally inflationary spacetime given by the metric \eqref{BubbleMetric2}, while $K_0 = \frac{2}{r}$ is the trace of the extrinsic curvature of $\mathbb{S}^2_r$ embedded in Minkowski spacetime \cite{BY:1993mass, Poisson:2004tool}. Explicitly, $K_g$ is computed as the divergence of the outward-pointing normal vector $\varsigma^a$ to $\mathbb{S}^2_r$ \cite{Poisson:2004tool}, i.e.
\begin{equation}
K_g = \nabla_a \varsigma^a.
\end{equation}

\noindent More explicitly, this gives a BY mass \cite{Mathematica}
\begin{equation}
M_\mathrm{BY} = r \left( e^{-2f} -1 \right) + \frac{r^2}{2} e^{-2f}\,(\partial_r f)
\end{equation}

\noindent for the local inflation bubble. We observe that both $M_\mathrm{MSH}$ and $M_\mathrm{BY}$ vanish exactly, whenever $f\in C_0^\infty(\mathbb{R}\times\mathbb{R}_+)$ and both of its derivatives $\partial_t f$ and $\partial_r f$ vanish simultaneously. In other words, there is no quasi-local gravitational energy, wherever the bubble geometry converges towards flat spacetime geometry, which is precisely the case outside of $\mathrm{supp}(f)$, as well as on possible plateaus, where both $\partial_t f$ and $\partial_r f$ are zero, but the scale factor remains larger than unity, i.e. $e^f > 1$. \\

In particular, recalling that the limit $r \rightarrow \infty$ of the BY mass recovers the Arnowitt-Deser-Misner (ADM) mass \cite{Poisson:2004tool}, see  also \cite{ADM:1961mass}, it follows immediately that
\begin{equation}
\label{ADMMass}
M_\mathrm{ADM} = \lim_{r\rightarrow\infty} M_\mathrm{BY} = 0,
\end{equation}

\noindent since $f\in C_0^\infty(\mathbb{R}\times\mathbb{R}_+)$ and hence $f=0$ for sufficiently large radius $r$. This confirms that the spacetime carries no net gravitational energy in the asymptotic region, while still exhibiting nontrivial quasi-local energy content within the local inflation bubble. \\

In addition, we  find that the spacetime does not radiate gravitational energy in the form of gravitational waves, which is proven  in the following way, using the Newman-Penrose (NP) formalism, see \cite{NP:1962f}. Consider the null tetrad $(l^a, n^a, m^a, \overline{m}^a)$, consisting of two real, future-directed null vector fields
\begin{align}
    \label{Outgoing}
    l^a &= \frac{1}{\sqrt{2}}\left(\frac{\partial}{\partial t} \right)^a + \frac{e^{-f}}{\sqrt{2}}\left(\frac{\partial}{\partial r} \right)^a, \\
    n^a &= \frac{1}{\sqrt{2}}\left(\frac{\partial}{\partial t} \right)^a - \frac{e^{-f}}{\sqrt{2}}\left(\frac{\partial}{\partial r} \right)^a,
    \label{Ingoing}
\end{align}

\noindent which satisfy $l^a n_a = -1$, and two complex null vectors
\begin{align}
    m^a &= \frac{e^{-f}}{\sqrt{2} \,r}\left(\frac{\partial}{\partial\vartheta} \right)^a + \frac{ie^{-f}}{\sqrt{2}\, r\sin\vartheta}\left(\frac{\partial}{\partial \varphi} \right)^a, \\
    \overline{m}^a &= \frac{e^{-f}}{\sqrt{2} \,r}\left(\frac{\partial}{\partial\vartheta} \right)^a - \frac{ie^{-f}}{\sqrt{2}\, r\sin\vartheta}\left(\frac{\partial}{\partial \varphi} \right)^a,
\end{align}

\noindent similarly satisfying $m^a \overline{m}_a = 1$. To analyze the presence of outgoing gravitational radiation in the spacetime \eqref{BubbleMetric}, we consider the NP scalar $\Psi_4$, which characterizes such radiation in asymptotically flat spacetimes \cite{BR:2016num}. This scalar quantity is defined in terms of the Weyl tensor $W_{abcd}$ as
\begin{equation}
\Psi_4 = W_{abcd} \, n^a \overline{m}^b n^c \overline{m}^d,
\end{equation}

\noindent where the Weyl tensor is defined in $(3+1)$ spacetime dimesions by \cite{Poisson:2004tool}
\begin{equation}
W_{abcd} = R_{abcd} - \left( g_{a[c} R_{d]b} - g_{b[c} R_{d]a} \right) + \frac{1}{3} R \, g_{a[c} g_{d]b},
\end{equation}

\noindent where $R_{abcd}$ is the Riemann tensor, $R_{ab}$ the Ricci tensor, $R$ the Ricci scalar, and $[\cdots]$ denotes antisymmetrization of indices. Recalling that the metric \eqref{BubbleMetric} is indeed asymptotically flat, since it coincides exactly with Minkowski spacetime outside the support of $f$, a direct computation for the case at hand shows that
\begin{equation}
\Psi_4 = 0.
\end{equation}
This equality confirms the absence of outgoing gravitational radiation in this geometric configuration. We are therefore dealing with a purely local construction that preserves the idealized properties of Minkowski spacetime at large scales. \\

Beyond the question of gravitational radiation, the tetrad $(l^a, n^a, m^a, \overline{m}^a)$ can also be used to investigate the formation of trapped surfaces, see \cite{Faraoni:2015horizons}, induced by the local inflation bubble. Hence, we compute the expansion scalars $\theta_l$ and $\theta_n$ corresponding to the outgoing and ingoing null congruences, see \cite{Poisson:2004tool}. A direct computation gives us \cite{Mathematica}
\begin{align}
\label{LExpansion}
\theta_l &= \nabla_a \, l^a = \sqrt{2} \left( \frac{3}{2} \, (\partial_t f) + e^{-f} \left( (\partial_r f) + \frac{1}{r} \right) \right), \\
\theta_n &= \nabla_a \, n^a = \sqrt{2} \left( \frac{3}{2} \, (\partial_t f) - e^{-f} \left( (\partial_r f) + \frac{1}{r} \right) \right).
\label{NExpansion}
\end{align}

\noindent These expansion scalars may vanish at specific radii $r > 0$, depending on the detailed form of the compactly supported function $f \in C_0^\infty(\mathbb{R}\times\mathbb{R}_+)$.
A bifurcating marginal surface emerges in the special case where $\partial_t f \to 0$,  either precisely on a plateau or approximately during periods of sufficiently slow inflation; this describes a unique geometric configuration in which both expansions simultaneously vanish, that is, $\theta_l = 0 = \theta_n$.   At such a surface, neither the outgoing nor the ingoing null congruence expands or contracts, and light remains (locally) stationary. More generally, however, any trapped surface, trapping horizon, or bifurcating marginal surface would necessarily be relatively short-lived, confined to the support of $f$, since for large values of $ t$ or $r$, the expansion scalars approach their Minkowski limits
\begin{align}
\theta_l &\longrightarrow \frac{\sqrt{2}}{r} > 0, \\
\theta_n &\longrightarrow -\frac{\sqrt{2}}{r} < 0.
\end{align}

\noindent We refer to Section \ref{SecEstimates} for a more detailed physical discussion of trapped surfaces and bifurcating marginal surfaces within local inflation bubbles, under the concrete physical assumptions and length scales involved. \\

Naturally, assuming the validity of general relativity and the Einstein field equations, we need to find the specific form of matter required to sustain the geometry described by \eqref{BubbleMetric2}. To this end, we compute the Einstein tensor explicitly, and obtain the nonvanishing components
\begin{align}
G_{ t t} &= 3(\partial_t f)^2 - \frac{e^{-2f}}{2r}\left( 8(\partial_r f) + 2r\,(\partial_r f)^2 + 4r(\partial_r^2) f\right), \\
G_{ t r} &= -2(\partial_t\partial_r f), \\
G_{rr} &= \frac{2}{r}(\partial_r f) + (\partial_r f)^2 - e^{2f}\left(3\,(\partial_t f)^2 + 2(\partial_t^2 f)\right), \\
G_{\vartheta\vartheta} &= r(\partial_r f) + r^2(\partial_r^2 f) - r^2\,e^{2f}\left( 3\,(\partial_t f)^2 + 2(\partial_t^2 f) \right), \\
G_{\varphi\varphi} &= G_{\vartheta\vartheta} \, \sin^2\vartheta.
\end{align}

\noindent Note that the corresponding energy momentum tensor $T_{ab} = \tfrac{1}{8\pi} G_{ab}$ does not \emph{a priori} describe any specific classical or semiclassical matter model. It rather needs to be understood as an effective source term that quantifies the (exotic) matter that is necessary to produce a local inflation bubble. However, we remark that in either of the special cases $\partial_t f \approx 0$ or $\partial_r f \approx 0$, that $T_{ab}$ (approximately) takes the form of an ideal fluid, since the component $T_{ t r}$ vanishes. \\

Moreover, using the spacetime's spherical symmetry, we may describe a general nonrotating timelike observer by the normalized vector field
\begin{equation}
u^a = \cosh\chi \, \left( \frac{\partial}{\partial  t} \right)^a +  e^{-f} \sinh\chi \, \left( \frac{\partial}{\partial r} \right)^a,
\end{equation}

\noindent which is parametrized by a centrifugal rapidity parameter $\chi \in \mathbb{R}$ and fulfills $u^a u_a = -1$. This allows us to express the WEC $T_{ab} u^a u^b \geq 0$ \cite{MV:2017ec,KS:2020ec} for the energy-momentum tensor $T_{ab}$ in terms of the function $f$. An explicit computation then yields the condition
\begin{align}
3 r\, e^{2f} (\partial_t f)^2 &\geq (\partial_r f) \left(3 + \cosh (2\chi)\right) + r(\partial_r f)^2 + 2r\, \cosh^2\chi\, (\partial_r^2 f) \nonumber \\
&\quad\; + 2r\, e^f \sinh(2\chi)( \partial_t \partial_r f )+ 2r\, e^{2f} \sinh^2\chi\,(\partial_r^2 f),
\label{WEC}
\end{align}

\noindent that must be fulfilled for all $( t, r) \in \mathbb{R} \times \mathbb{R}_+$ and all $\chi \in \mathbb{R}$ for the WEC to be satisfied. We observe, however, that \eqref{WEC} is generally not fulfilled everywhere, even for static observers, i.e. for $\chi = 0$. Hence, the energy density measured along timelike worldlines is not everywhere positive, and any matter sourcing the geometry of a local inflation bubble must necessarily violate the WEC. \\ 

Similarly, we find that the null energy condition (NEC) is generically violated by our construction. This follows from the focusing theorem for null geodesic congruences, which states that for hypersurface orthogonal null geodesics the expansion scalar $\theta$ cannot increase along the affine parameter whenever the NEC is satisfied \cite{Poisson:2004tool}. Since our construction is specifically designed to induce a local expansion, including a defocusing of null geodesics, it straightforwardly follows that the spacetime must necessarily violate the NEC. More generally, we also expect the model to violate averaged energy conditions, including the (achronal) ANEC, due to its continued reliance on exotic stress-energy contributions. A definitive assessment of the (achronal) ANEC, however, would require a considerably more extensive analysis along complete null geodesics, and since traversable wormhole geometries inherently require exotic matter and are accordingly expected to violate energy conditions, such an analysis lies beyond the present scope. \\

Nevertheless, any negative expressions in the matter sector are bounded from below by construction, since the local inflation bubble is compactly supported via the function $f\in C_0^\infty(\mathbb{R}\times\mathbb{R}_+)$ and transitions smoothly into Minkowski outside the bubble. Consequently, all curvature quantities, and hence all components\footnote{In particular, this holds for $\rho=\left.T_{ab}u^{a}u^{b}\right|_{\chi=0}$.} of $T_{ab}$, are bounded functions in $ t$ and $r$. This implies that there exists a constant $K>0$ such that the energy density $\rho$ is bounded from below by $\rho( t,r)\ge -K$ for all $ t$ and $r$. This uniform lower bound is an essential feature of the local inflation bubble, since it rules out arbitrarily large negative energies and mirrors the idea of quantum energy inequalities\footnote{To meaningfully compare the pointwise lower bound $\rho \geq -K$ with lower bounds established by QEIs, one must first average the classical (exotic) energy density along a suitable worldline. This is because QEIs do not impose pointwise constraints on the energy density, but rather provide lower bounds only after appropriate averaging.}, which temporarily allow for negative energy densities within a given time window.  \\

To quantify the energetic cost of this mechanism, we consider the total energy $E_\mathrm{stat}$ contained in a constant-time surface, as measured by a static observer with four-velocity $u^{a}|_{\chi=0} = (\tfrac{\partial}{\partial t})^{a}$. Explicitly, we obtain
\begin{align}
E_\mathrm{stat}( t) &= \int_{\Sigma_t} \left. T_{ab}\, u^a u^b \right\vert_{\chi=0} \, d\mathrm{vol}_{\Sigma_t} = \int_{\Sigma_t} T_{tt} \, d\mathrm{vol}_{\Sigma_t} \nonumber \\
&= \frac{1}{8\pi} \int_0^\infty \int_0^\pi \int_0^{2\pi} \left( 3(\partial_t f)^2 - e^{-2f} \left( \frac{4}{r}(\partial_r f) + (\partial_r f)^2 + 2(\partial_r^2 f) \right) \right) e^{3f} r^2 \sin\vartheta \, dr \, d\vartheta \, d\varphi \nonumber \\
&= \frac{1}{2} \int_{\mathrm{supp}(f)} \left( 3r^2 e^{3f} (\partial_t f)^2 - 4r e^f (\partial_r f) - r^2 e^f (\partial_r f)^2 - 2r^2 e^f (\partial_r^2 f) \right) \, dr
\end{align}

\noindent for which we can use integrate by parts, i.e.
\begin{equation}
 - \frac{1}{2} \int_{\mathrm{supp}(f)} r^2 \left( e^f (\partial_r f)^2 + e^f \partial_r^2 f \right) \, dr=\int_{\mathrm{supp}(f)} r e^f (\partial_r f) \, dr,  
\end{equation}

\noindent in order to show that
\begin{equation} \boxed{
E_\mathrm{stat}( t)  =     \frac{1}{2} \int_{\mathrm{supp}(f)} \left( 3 e^{3f} (\partial_t f)^2 + e^f (\partial_r f)^2 \right) r^2 \, dr \geq 0. }
\label{PositiveEnergy}
\end{equation}

\noindent This result demonstrates that, despite pointwise violations of energy conditions, the integrated energy within a given equal-time slice, in the foliation associated to static worldlines, is manifestly nonnegative. Consequently, although the geometry certainly requires exotic matter contributions locally, it can still appear relatively well-behaved from a macroscopic perspective. \\

We point out, however, that this notion of energy explicitly depends on the specific choice of observer and the associated foliation of the spacetime by equal-time surfaces, so that the integral \eqref{PositiveEnergy} generally yields different expressions for different values of the parameter $\chi$. Consequently, $E_\mathrm{stat}$ is conceptually distinct from the previously computed global ADM mass \eqref{ADMMass} and the quasi-local Brown-York mass \eqref{BrownYorkMass}, which provide invariant geometric definitions for the gravitational energy. In contrast, $E_\mathrm{stat}(t)$ quantifies the integrated local energy density $T_{tt}$ in the reference frame where the local inflation bubble is at rest. Physically, this corresponds to the total energy that must be supplied at any given coordinate time $t$ to sustain the localized geometric deformation. \\

More generally, the volume-integrals $\int_{\Sigma_t}\rho\,d\mathrm{vol}_{\Sigma_t}$ and $\int_{\Sigma_t}(\rho+p_r)\,d\mathrm{vol}_{\Sigma_t}$ are frequently employed as global quantifiers of (averaged and pointwise) energy condition violations in ideal fluid settings with the identifications $\rho=T_{tt}$ and $p_r=T_{rr}$, see \cite{VKD:2003prl,VKD:2004wq}. For the local inflation bubble, $\int_{\Sigma_t}\rho\,d\mathrm{vol}_{\Sigma_t}$ is intrinsically nonnegative by construction, whereas an explicit evaluation \cite{Mathematica} shows that $\int_{\Sigma_t}(\rho+p_r)\,d\mathrm{vol}_{\Sigma_t}$ can indeed become negative, sometimes substantially so. This indicates that a negative radial pressure is the primary source of the aforementioned energy condition violations. As explained above, the positive integral $\int_{\Sigma_t}\rho\,d\mathrm{vol}_{\Sigma_t}$ describes the observer-dependent total energy on an equal-time slice $\Sigma_t$ and does not take the negative stress contributions into account that are captured by $\int_{\Sigma_t}(\rho+p_r)\,d\mathrm{vol}_{\Sigma_t}$. As an interpretational caveat, however, we point out that the identifications $\rho=T_{tt}$ and $p_r=T_{rr}$ only hold exactly for ideal fluids and need not apply to more general exotic matter sources.

\section{Concrete Model and Energy Estimates}\label{SecEstimates}
An explicit model for the local inflation of a spacetime region centered around the origin is constructed using the standard smooth bump function
\begin{equation}
b(x) = \begin{cases} 
    e^{-\frac{1}{1 - x^2}}, & \text{for } \vert x \vert < 1, \\
    0, & \text{otherwise},
\end{cases}
\end{equation}

\noindent which is has compact support in the interval $[-1,1]$ and is an element of the space $C_0^\infty(\mathbb{R})$. Based on the function $b$, we define the compactly supported function $f \in C_0^\infty(\mathbb{R}\times\mathbb{R}_+)$ by

\begin{equation}
\label{DoubleBump}
f( t, r) = L^2 \, b\left(\frac{ t}{\Delta t}\right) \, b\left(\frac{r}{\Delta r}\right),
\end{equation}

\noindent where $\Delta  t, \Delta r > 0$ determine the temporal and radial extent of the spacetime region, while $L \in \mathbb{R}$ sets the amplitude of the inflation via $f(0,0) = \frac{L^2}{e^2}$. In particular, the partial derivatives of $f$ scale as $\partial_y f \sim \mathcal{O}\left(\frac{f}{\Delta y}\right)$ for $y =  t$ or $y = r$, reflecting the controlled sharpness of the bump profile in each direction. This choice of $f\in C_0^\infty(\mathbb{R}\times\mathbb{R}_+)$ enables us to model spatially localized expansion over finite time, whose geometric and energetic properties can be analyzed in full detail under variation of the model parameters $L$, $\Delta t$, and $\Delta r$. \\

\begin{figure}[h!]
\centering
\includegraphics[width=0.44\textwidth]{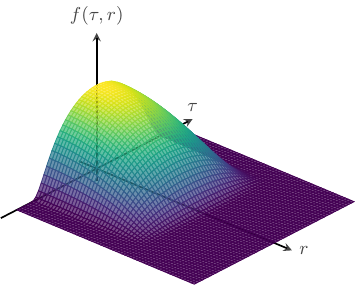}
\caption{\justifying \small {Schematic plot of the double bump function $f( t, r)$ defined in equation \eqref{DoubleBump}, highlighting its compact support and smooth decay in both temporal and radial directions. For illustrative purposes, the parameters have been chosen as $\Delta  t = 2$, $\Delta r = 0.1$, and $L=0.2$}, which does \emph{not} correspond to the physically viable assumptions discussed in Section \ref{SecEstimates}.}
\label{BumpPlot}
\end{figure}

\begin{figure}[h!]
\centering
\includegraphics[width=0.8\textwidth]{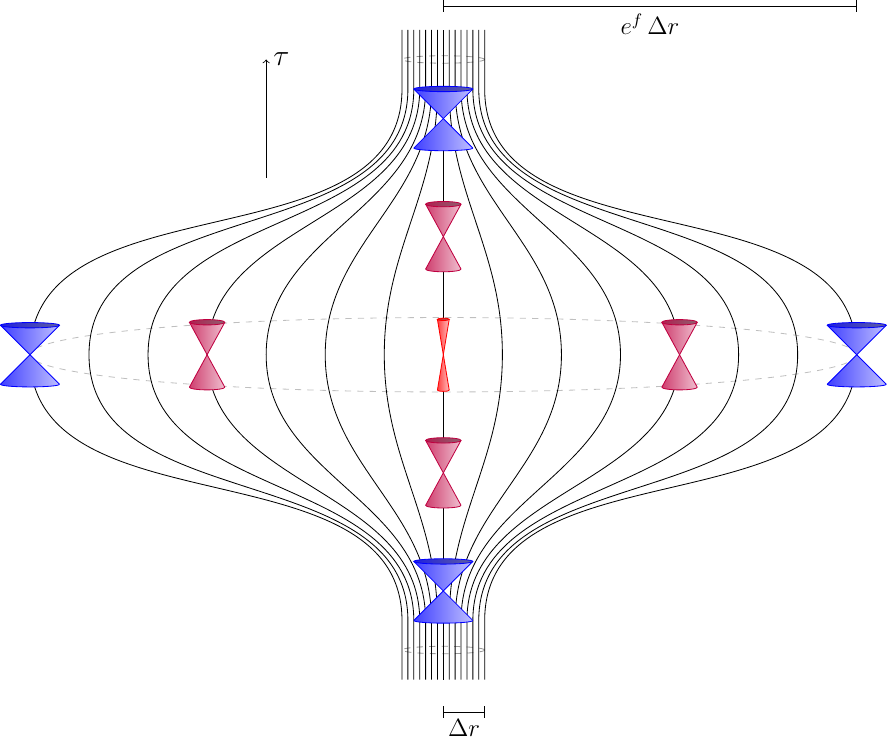}
\caption{\justifying \small{Schematic visualization of the local inflation and contraction of the spatial region $\left\lbrace (r,\Omega)\in\Sigma \; \vert \; r<\Delta r \right\rbrace$ over time $ t$, as described by the metric \eqref{BubbleMetric} with the function $f( t,r)$ given by \eqref{DoubleBump}. Shown are nonrotating worldlines of constant $r < \Delta r$, as well as local lightcones at various points in spacetime. In the central region (red), where the function $f( t, r)$ reaches its maximum, the lightcones become noticeably steeper and narrower, reflecting the local modification of causal structure induced by the inflationary stretching, see expressions \eqref{Outgoing} and \eqref{Ingoing}. In contrast, in regions where $f$ is close to zero (blue), the lightcones approach the standard 45-degree angles of lightcones in Minkowski spacetime.}}
\label{BubbleDiagram}
\end{figure}

Substituting the function \eqref{DoubleBump} into the spacetime metric \eqref{BubbleMetric} yields a geometry that remains identical to Minkowski spacetime with the exception of the compact region,
\begin{equation}
\mathrm{supp}(f) = \left\lbrace \left. ( t,r,\Omega)\in\mathbb{R}^4 \; \right\vert\; \vert  t \vert < \Delta t \; , \; r< \Delta r \right\rbrace,
\end{equation}

\noindent where the geometry undergoes a rapid but smooth and finite inflation, as well as a subsequent contraction. More specifically, the region expands rapidly during the interval $-\Delta t <  t < 0$, with the rate governed by the parameter $L$, and contracts back to flat spacetime over the interval $0 <  t < \Delta t$. This transient deviation from flatness is illustrated schematically in Figures \ref{BumpPlot} and \ref{BubbleDiagram}. \\

\begin{figure}[h!]
\centering
\includegraphics[width=0.9\textwidth]{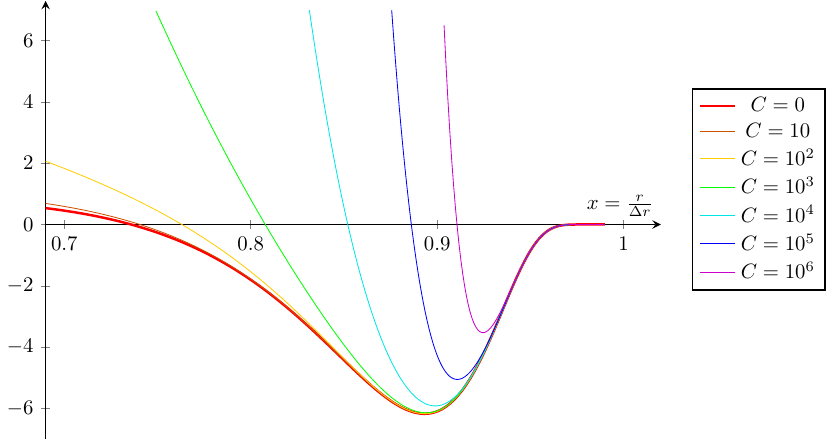}
\caption{\justifying \small{Numerical plot of the spatial regions where inequality \eqref{StationaryEnergyDensityINEQ} may be violated, i.e. where the energy density observed by a static observer becomes locally negative. For $0 < x < 0.7$, the inequality is satisfied in all cases shown. The $y$-axis represents the left hand side of inequality \eqref{StationaryEnergyDensityINEQ} evaluated at the fixed coordinate time 
$ t = \pm\, \Delta t \, \sqrt{1 - \frac{1}{2 \ln (L)}}$, 
valid for $L \geq \sqrt{e}$, which is chosen for computational simplicity. At this choice of $t$, we effectively parametrize the order of magnitude of the first (positive) term in inequality \eqref{StationaryEnergyDensityINEQ} by the parameter $C := \left(\frac{\Delta r}{\Delta  t}\right)^216\,\ln(L)^4 \left(1 - \frac{1}{2 \ln (L)}\right)$. Note that for more general choices of $ t$, the magnitude of the first term in inequality \eqref{StationaryEnergyDensityINEQ} depends sensitively on both $ t$ itself and the inflation amplitude $L$. Different curves in the plot correspond to different values of $C$, which increases for shorter time intervals $\Delta t$, larger bubble radii $\Delta r$, or larger inflation amplitudes $L$. We observe that increasing $C$ suppresses the extent of violations of positive energy density. However, the physically relevant regimes corresponding the values given in Table \ref{TableParameterRegimes} rather lead to $C \approx 0$, for which the violations of inequality \eqref{StationaryEnergyDensityINEQ} near the boundary $r \lesssim \Delta r$ are nonnegligible, as depicted by the red curve.}}
\label{INEQPlot}
\end{figure}

Moreover, let us examine the positivity of the energy density, as observed by a static observer, corresponding to inequality \eqref{WEC} with $\chi = 0$, which explicitly yields the condition
\begin{align}
\frac{3(\Delta r)^2}{4(\Delta t)^2} \, x \,e^{2f}\frac{\tau^2}{\left( \tau^2 - 1 \right)^4} f^2 + \frac{2x}{\left( x^2 - 1 \right)^2} f - \frac{x^3}{\left(x^2 - 1 \right)^4} f^2 - \frac{x \left( 3x^4 - 1 \right)}{\left( x^2 - 1 \right)^4} f^2 \geq 0,
\label{StationaryEnergyDensityINEQ}
\end{align}

\noindent where we have used the shorthand notation $\tau = \tfrac{t}{\Delta t}$ and $x = \tfrac{r}{\Delta r}$. A straightforward numerical analysis of inequality \eqref{StationaryEnergyDensityINEQ} for specific values of $ t$ indicates that violations typically occur at relatively large values of $r \in \mathrm{supp}(f)$, see Figure \ref{INEQPlot}. However, the magnitude and spatial extent of these violations are substantially suppressed by choosing smaller time intervals $\Delta t$, larger maximal radii $\Delta r$, or larger inflation amplitudes $L$, as illustrated Figure \ref{INEQPlot}. \\

Let us now consider specific assumptions for the orders of magnitude of the parameters $\Delta  t$, $\Delta r$, and $L$, which characterize the temporal duration, spatial extent, and amplitude of the local inflation, respectively. We emphasize that the following considerations provide heuristic order-of-magnitude estimates that explicitly rely on the assumption that the compactly supported function takes the specific form \eqref{DoubleBump}. More precise estimates or more general profiles, such as Expression \eqref{SuperpositionBump} discussed below, the computations are expected to become significantly more involved. \\

We argue that in order to make the physical effects of quantum geometry visible, the spatial scale $\Delta r$ should correspond to a regime where quantum gravitational phenomena, such as a noncommutative geometry, are expected to become significant. This suggests a characteristic length scale of lower order multiples of the Planck length, and we adopt the estimate $\Delta r \approx 10^2\, \ell_P$. In contrast, the temporal scale $\Delta  t$ should be relatively macroscopic, since the central idea is to smoothly amplify quantum-scale phenomena to observable distances, over observable time intervals. As an illustrative choice, we consider $\Delta  t$ to be of the order of a few seconds, corresponding to $\Delta  t \approx 10^{44}\, t_P$ in Planck times. At last, the amplitude parameter $L$ must be chosen such that the initially microscopic spatial region of radius $\Delta r$ undergoes sufficient inflation to reach macroscopic dimensions, ideally on the order of a few metres. To this end, we assume that the maximal value of the scale factor $e^f$ should be approximately of the order of magnitude $10^{33}$, which implies $f(0,0) \approx 76$. From the definition \eqref{DoubleBump}, this yields an estimate of $L \approx \sqrt{76}\, e \approx 23.7$. \\

Under these numerical assumptions, the parameter $C$ defined underneath Figure \ref{INEQPlot} becomes effectively zero, i.e. $C \approx 0$. As a result, local violations of inequality \eqref{StationaryEnergyDensityINEQ} arise in the form of nonnegligible negative energy densities, which remain bounded from below. Most importantly, however, the total energy per constant-time slice along a static worldline remains positive at all times, as shown in equation \eqref{PositiveEnergy}. Therefore, let us estimate the maximal amount of energy contained in the spacetime configuration \eqref{BubbleMetric} for the specific model \eqref{DoubleBump}, under the assumed parameter values. Expressing \eqref{PositiveEnergy} in Planck units, we note that the terms in the integrand scale as
\begin{align}
e^{3f} (\partial_t f)^2 &\sim \left(10^{33}\right)^3 \left(10^{44}\right)^{-2} = 10^{11} \\ 
e^f (\partial_r f)^2 &\sim 10^{33} \left(10^2\right)^{-2} = 10^{29}.  
\end{align}

\noindent Integrating over the interval $[0, \Delta r]$, including the radial factor $r^2$, we obtain a maximal energy contained in an equal-time surface associated to a static observer of the order
\begin{equation}
E_\mathrm{stat}( t) \lesssim \mathcal{O}(10^{35}).
\end{equation}

\noindent When restoring natural constants and converting to CGS units, this corresponds to approximately $10^{51}\, \mathrm{erg}$, which is comparable to the energy released in a typical supernova explosion, see \cite{HCBF:1997,Hartmann:1999sn,Weinberg:2006bethe}. While such an energy production would certainly lie within the capabilities of a hypothetical Kardashev type III civilization, see \cite{Kardashev:1964civ}, it remains far beyond the technological and energetic reach of humanity for the foreseeable future. Therefore, let us turn to a more conservative energy estimate, grounded in the limits of presently available technology. \\

Most prominently, the shortest currently measurable time intervals lie in the range of a few attoseconds \cite{KGFAAS:2010as}, which translates to approximately $\Delta  t \approx 10^{27}\, t_P$ in Planck times. Similarly, the highest precision in length measurements is achieved by gravitational wave observatories, such as LIGO, which are sensitive to displacements on the order of attometres \cite{Aetal:2016gw}. This implies that the minimal inflation amplitude required to render any effects of quantum spacetime measurable must be at least of the same order, i.e. $e^{f(0,0)} \approx 10^{17}$, corresponding to $L \approx 17$. Under these assumptions, the parameter $C$ (see Figure \ref{INEQPlot}) increases by several orders of magnitude compared to the previous case, yet remains well below unity. Consequently, the qualitative behaviour of the energy density, including the persistence of local negative energy regions, is not significantly altered when compared to the previously considered scenario. However, the total energy content of an equal-time surface associated to a static worldline changes substantially, as the following estimates show. Reconsidering expression \eqref{PositiveEnergy}, we find that the terms now (maximally) scale as
\begin{align}
e^{3f}(\partial_t f)^2 &\sim \left( 10^{17} \right)^3\left( 10^{27} \right)^{-2} = 10^{-3} \\
e^f (\partial_r f)^2 &\sim 10^{17} \left( 10^2 \right)^{-2} = 10^{13}.
\end{align}

\noindent In analogy to the previous scenario, the resulting energy integral is therefore at most of the order of magnitude
\begin{equation}
E_\mathrm{stat}( t) \lesssim \mathcal{O}\left(10^{19}\right).
\end{equation}

\noindent Converting to SI units, this corresponds to approximately $10^{28}\,\mathrm{J}$, which still exceeds the current global annual energy consumption by several orders of magnitude, see \cite{RRR:2020energy}, and even surpasses the total estimated reserves of nonrenewable energy resources worldwide \cite{BP:2010energy}. If these estimates do not provide a sufficient incentive to accelerate the development of renewable energy technologies, such energy budgets likely remain exclusive to type II civilizations on the Kardashev scale \cite{Kardashev:1964civ}. \\

\begin{table}[h]
\centering
\begin{tabular}{|l|c|c|c|c|}
\hline
\textbf{Expansion Scale} & $\Delta t$ [$t_P$] & $\Delta r$ [$l_P$] & $L$ & $E_{\text{stat}}$ \\ \hline
Macroscopic & $10^{44}$ & $10^2$ & $23.7$ & $\lesssim \mathcal{O}(10^{35} E_P) \approx 10^{51}$ erg \\ \hline
Attoscale & $10^{27}$ & $10^2$ & $17.0$ & $\lesssim \mathcal{O}(10^{19} E_P) \approx 10^{28}$ J \\ \hline
\end{tabular}
\caption{\justifying {\small Summary of the specific parameter choices and corresponding energy estimates for the different physical expansion scales discussed in Section \ref{SecEstimates}.}}
\label{TableParameterRegimes}
\end{table}

Finally, we address the question of whether the locally inflated region, and the amplified quantum-induced structures within it, are in principle accessible or at least observable. Recalling the geometric quantities discussed in Section \ref{GeometricConstruction}, together with the specific model given by the function \eqref{DoubleBump}, we now examine the outgoing and ingoing null vector fields introduced in equations \eqref{Outgoing} and \eqref{Ingoing}. Near the point of maximal expansion, i.e. $( t, r) = (0, 0)$, the radial propagation of null geodesics, and consequently that of all timelike observers, is strongly suppressed due to the exponential factor $e^{-f}$, which, depending on the parameter choice, can take values as small as $10^{-33}$ to $10^{-17}$. Physically, this means that near the central region of the bubble, all causal worldlines become nearly parallel, so that any attempt to enter or exit the region is drastically slowed down during the inflationary phase. Nevertheless, access and escape remain, in principle, possible, with the only requirement that the duration of the inflationary phase $\Delta t$ exceeds the corresponding travel time of the entering or escaping signal or observer. In cases of extreme suppression, this travel time may become so long that the endeavour becomes rather impractical, even if it is not prohibited by the geometry. At the very latest, once the bubble has contracted back to the flat Minkowski geometry, signals or observers can freely enter or exit the formerly inflated region, so that any events that occurred within the bubble are, in principle, detectable by observers in its causal future.  \\

This behaviour is also reflected in the properties of geodesic null congruences and trapped surfaces. Evaluating the expansion scalars \eqref{LExpansion} and \eqref{NExpansion}, we find that both become very small in the central region of the inflation bubble, where the terms involving $\partial_t f$ are suppressed by several orders of magnitude more than the remaining contributions. Consequently, the expansions satisfy the approximate relation $\theta_l \approx -\theta_n$, implying that any surface with vanishing outgoing expansion, i.e. $\theta_l = 0$, also has an (almost) vanishing ingoing expansion, $\theta_n \approx 0$. Instead of forming a regular trapping horizon, the geometry thus produces approximately bifurcating marginal surfaces, which is geometrically similar to a static wormhole throat \cite{HV:1998dw,MHC:2009cw}, even though no conventional wormhole geometry is present thus far. Physically, this corresponds to a state in which both outgoing and ingoing light rays are neither diverging nor converging, consistent with the depiction of steep lightcones and nearly parallel causal curves near the bubble’s center, as depicted in Figure \ref{BubbleDiagram}.

\section{Local Inflation of Wormholes}\label{WormholeSection}
The final step in this work is to embed a wormhole (see the corresponding metric given in Equation \eqref{WormholeMetric}) into a local inflation bubble. Therefore, we assume that the throat radius $s_0$ is much smaller than the radial extent of the support of the function $f \in C_0^\infty(\mathbb{R}^2)$ generating the inflation. Denoting by $\Delta l$ the proper-length radius of $\mathrm{supp}(f)$, this requirement is expressed by $s_0 \ll \Delta l$, ensuring that the wormhole lies entirely within the inflated domain. In analogy to the model proposed in \cite{Roman:1993wh}, the resulting combined line element then takes the form

\begin{equation}
\boxed{ds^2 = -dt^2 + e^{2f(t,l)} \left( dl^2 + \left( s_0^2 + l^2 \right)\, d\Omega^2 \right),}
\label{EmbeddedWormholeMetric2}
\end{equation}

\noindent describing a bubble of finite spatial and temporal extent that undergoes rapid inflation followed by contraction, with the wormhole fully enclosed. In particular, the wormhole throat radius, initially at $s_0$, inflates as $e^{f} s_0$ during the inflationary phase. Outside of $\mathrm{supp}(f)$, the metric reduces exactly to the original wormhole geometry \eqref{WormholeMetric}, which is already nearly flat under the assumption $s_0 \ll \Delta l$, while the wormhole’s characteristic structure is entirely contained within, and dynamically affected by, the local inflation bubble. Again, we refer to appendix \ref{AppendixCurvatureResults} for the exact results for the curvature quantities associated with the metric \eqref{EmbeddedWormholeMetric2}. \\

Analogously to the case of a pure local inflation bubble, as discussed in Section \ref{GeometricConstruction}, we are particularly interested in the energy density measured by static observers, corresponding to the temporal component of the stress-energy tensor. Given the metric \eqref{EmbeddedWormholeMetric2}, this component explicitly takes the form
\begin{equation}
\label{T00EmbeddedWormhole}\boxed{
T_{tt} = \frac{3}{8\pi}(\partial_t f)^2 - \frac{e^{-2f}}{8\pi} \left( \frac{s_0^2}{\left( s_0^2 + l^2 \right)^2} + \frac{4l}{s_0^2 + l^2}(\partial_l f) + (\partial_l f)^2 + 2\,(\partial_l^2 f) \right).}
\end{equation}

\noindent Note that the sign of $T_{tt}$ is generally undetermined, implying that the local energy density can be locally negative, although being bounded from below by definition, as in the previous case. Furthermore, just as the isolated inflation bubble and the Morris-Thorne wormhole individually violate the WEC and NEC, the combined system generally violates these pointwise energy conditions as well. More generally, violations of averaged energy conditions, including the (achronal) ANEC, are also expected, since such violations already occur for a Morris-Thorne wormhole, which necessarily relies on exotic matter, and the presence of the local inflation bubble is not expected to qualitatively change this behavior. \\

Using the volume element $\sqrt{-g} = e^{3f}\left( s_0^2 + l^2 \right)\,\sin\vartheta$, we compute the total energy of a constant-time slice corresponding to a static worldline as integral of $T_{tt}$ over a hypersurface of constant $t$, in order to obtain
\begin{align}
\int_{\Sigma_t} T_{tt} \, d\mathrm{vol}_{\Sigma_t} &= \frac{1}{2}\int_0^\infty \biggl( 3e^{3f} \left( s_0^2 + l^2 \right) (\partial_t f)^2 - e^f \frac{s_0^2}{s_0^2 + l^2} - 4l \, e^f (\partial_l f) \nonumber \\
&\hspace{2.5cm} - e^f \left( s_0^2 + l^2 \right) (\partial_l f)^2 - 2 e^f \left( s_0^2 + l^2 \right) (\partial_l^2 f) \biggl) dl \\
&= \frac{1}{2}\int_0^\infty \biggl( 3e^{3f} \left( s_0^2 + l^2 \right) (\partial_t f)^2 + l^2 e^f (\partial_l f)^2 - e^f \frac{s_0^2}{s_0^2 + l^2} \nonumber \\
&\hspace{4.9cm}- e^f s_0^2 \, (\partial_l f)^2 - 2 e^f s_0^2\, (\partial_l^2 f) \biggl) dl \\
&= E_\mathrm{stat}^0 (t) + \frac{s_0^2}{2} \int_0^\infty \left( 3e^{3f} (\partial_t f)^2 - e^f \, (\partial_l f)^2 - 2 e^f \, (\partial_l^2 f) - e^f \frac{1}{s_0^2 + l^2} \right) dl,
\label{InflatedWormholeEnergy}
\end{align}

\noindent where $E_\mathrm{stat}^0$ denotes the positive energy of the pure inflation bubble, given in \eqref{PositiveEnergy}, which is independent of the wormhole geometry. \\

A comment on the range of integration is in order before turning to the energetic analysis of the embedded wormhole. The proper radial coordinate $l \in \mathbb{R}$ covers the full two-ended Morris-Thorne geometry, with $l=0$ corresponding to the wormhole throat and the two asymptotically flat regions located at $l \to \pm\infty$. By construction, we choose the profile $f(t,l) \in C_0^\infty(\mathbb{R}\times\mathbb{R})$ to be symmetric under $l \leftrightarrow -l$, so that all curvature quantities and stress-energy components relevant to our energy estimates, most importantly the integrand in Expression \eqref{InflatedWormholeEnergy}, depend on $l$ only through $f$ and derivatives thereof, and are therefore likewise symmetric in $l$. Consequently, integrating over a single asymptotic region, $l \in [0,\infty)$, versus over the full two-ended geometry, $l \in \mathbb{R}$, differs only by an overall factor of $2$ and does not qualitatively change the results and their subsequent discussion. \\

We observe that the total energy per constant-time surface associated to a static observer thus not only contains the sum of $E_\mathrm{stat}^0$ and the negative energy of the wormhole (rescaled by a factor $e^{f}$ due to the local inflation), but also additional cross-terms arising from the interaction between the wormhole geometry and the local expansion. It is worth noting that the rescaled wormhole term is no longer compactly supported within $\mathrm{supp}(f)$. However, it remains finite, since we explicitly find \cite{Gradshteyn:1980int}
\begin{equation}
\label{ExplicitWormholeIntegral}
\int_{0}^{\infty} e^{f} \,\frac{1}{s_0^2 + l^2} \, dl
= \underbrace{\frac{e^{f}}{s_0} 
\left. \arctan\!\left( \frac{l}{s_0} \right) \right|_{0}^{\infty}}_{= \frac{\pi}{2 s_0} \; \forall \, f \in C_0^\infty(\mathbb{R}^2)}
- \frac{1}{s_0} \,\underbrace{\int_{0}^{\infty} \arctan\!\left( \frac{l}{s_0} \right) e^{f} (\partial_l f) \, dl}_{<\,\infty}\, < \infty \, ,
\end{equation}

\noindent which confirms that Expression \eqref{InflatedWormholeEnergy} remains well-defined, and most importantly bounded, even though its spatial support partially extends beyond $\mathrm{supp}(f)$. In contrast to the pure inflation bubble geometry, the total energy of a constant-time surface associated to a static observer is no longer positive in the combined system, and can, in fact, become negative for certain choices of $s_0 > 0$ and $f \in C_0^\infty(\mathbb{R}^2)$ at specific coordinate times $t$. Note, however, that not only the energy density, but also the total energy remain bounded from below for all possible configurations within the framework, as in the previous case. \\

In particular, we point out two different possibilities to render the energy density \eqref{T00EmbeddedWormhole} positive. The first is to relax the assumption that $f$ has compact support in $(t,l)$ and admit the de Sitter ansatz $f(t,l)=f(t)=H t$ with constant $H>0$, retrieving the known scenario of wormholes undergoing cosmic inflation, which is discussed in \cite{Roman:1993wh}. Consistently, we find that the energy density of a traversable wormhole embedded in an expanding de Sitter universe takes the form
\begin{equation}
T_{tt}^\mathrm{dS} = \frac{3}{8\pi} H^2 - e^{-2Ht} \frac{s_0^2}{8\pi\,\left(s_0^2 + l^2\right)^2},
\end{equation}

\noindent which is bounded from below by
\begin{equation}
T_{tt}^\mathrm{dS} \geq \frac{3}{8\pi} H^2 - \frac{e^{-2Ht}}{8\pi\,s_0^2},
\label{dSWormholeLowerBound}
\end{equation}

\noindent for all $l\in\mathbb{R}$. This expression becomes positive, if and only if
\begin{equation}
3H^2 \geq \frac{e^{-2Ht}}{s_0^2},
\end{equation}

\noindent which is, for $t>0$, fulfilled if \cite{Mathematica}
\begin{equation}
\label{HubbleParameterBound}
H \geq \frac{1}{t} W_0\left( \frac{t}{\sqrt{3}\,s_0} \right),
\end{equation}

\noindent where $W_0$ denotes the Lambert $W$ function. Observing that the right-hand side of \eqref{HubbleParameterBound} tends to zero as $t\to\infty$ \cite{Mathematica}, the energy density $T_{tt}^\mathrm{dS}$ eventually becomes positive, since for each $H>0$ there exists some $t_\ast$ such that $T_{tt}^\mathrm{dS}(t,l)>0$ for all $t \geq t_\ast$ and $l\in\mathbb{R}$. \\

\begin{figure}[h!]
\centering
\includegraphics[width=0.82\textwidth]{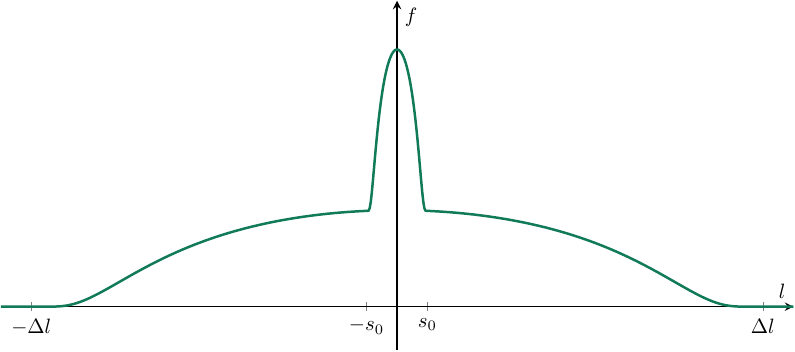}
\caption{\justifying \small{Sketch of a smooth, compactly supported function $f(l)$ that is constructed as the superposition of two individual bump functions, such that the combined profile is both sufficiently extended in $l$-direction and sufficiently sharply peaked around $l=0$. For illustrative purposes, the parameters have been chosen as $\Delta l_1 = 12$ and $\Delta l_2 = 1$.}}
\label{Illustration}
\end{figure}

Secondly, within the local inflation bubble framework, given by a compactly supported function $f\in C_0^\infty(\mathbb{R}^2)$, the positivity of the energy density at the throat of the embedded wormhole is ensured by requiring that $f$ is sufficiently sharply peaked at $l=0$. More precisely, we assume that $f$ possesses a strict local maximum at $l=0$, so that $\left.\partial_l f\right\vert_{l=0}=0$ and $\left.\partial_l^2 f\right\vert_{l=0}<0$. Under these specific assumptions, the energy density at the throat takes the form
\begin{equation}
\label{ThroatDensity}
T_{tt}^{l=0} = \frac{1}{8\pi}\left(3(\partial_t f)^2 -e^{-2f} \left( 2(\partial_l^2 f) + \frac{1}{s_0^2} \right) \right),
\end{equation}

\noindent which becomes manifestly positive if
\begin{equation}
\partial^2_l \Big\vert_{l=0} f < -\frac{1}{2\,s_0^2},
\label{SharpnessCondition}
\end{equation}

\noindent i.e. if the the profile of $f$ at the wormhole throat is sufficiently concave. Unfortunately, for functions of the form \eqref{DoubleBump}, the condition \eqref{SharpnessCondition} is in tension with the scale requirement $\Delta l \gg s_0$, since in this case $\left. \partial_l^2 \right\vert_{l=0} f = -\tfrac{f}{(\Delta l)^2} \gg -\tfrac{1}{2s_0^2}$. Nevertheless, it is generally possible to construct more involved compactly supported functions $f\in C_0^\infty(\mathbb{R}^2)$ that simultaneously satisfy both \eqref{SharpnessCondition} and $\Delta l \gg s_0$, e.g. by decoupling the throat curvature from the large-$l$ decay. The simplest example for this would be the superposition of two different bump functions of the form \eqref{DoubleBump} that fulfill both properties individually, as illustrated in Figure \ref{Illustration}. More precisely, such a function can be constructed as
\begin{equation}
\label{SuperpositionBump}
f(t,l) = f_1(t,l) + f_2(t,l),
\end{equation}

\noindent where $f_1$ is relatively flat bump function with $\Delta l_1 \gg s_0$, while $f_2$ is relatively sharpely peaked around $l=0$ with individual $l$-support significantly smaller than the $l$-support of $f_1$, i.e. $\Delta l_2 \ll \Delta l_2$. The second $l$-derivative of \eqref{SuperpositionBump} at the throat is then given by
\begin{equation}
\label{SecondSuperpositionDerivative}
   \partial^2_l \Big\vert_{l=0} f = - \frac{f_1}{(\Delta l_1)^2} - \frac{f_2}{(\Delta l_2)^2}.
\end{equation}

\noindent Choosing $\Delta l_2$ to be sufficiently small, such that Expression \eqref{SecondSuperpositionDerivative} fulfills the concavity condition \eqref{SharpnessCondition}, we observe that the energy density \eqref{ThroatDensity} becomes positive, while the scale requirement $\Delta l = \Delta l_1 \gg s_0$ is still satisfied. \\

Finally, we briefly repeat the energy estimates from Section \ref{SecEstimates} for an inflated Planck-scale wormhole with $s_0 \sim 1$. Using the specific assumptions from Table \ref{TableParameterRegimes}, we find that all additional negative energy contributions in Expression \eqref{InflatedWormholeEnergy} at the time of maximal expansion $t=t_\mathrm{max}$ are of subleading orders of magnitude compared to $E_\mathrm{stat}^0(t_\mathrm{max})$. Consequently, the overall energy estimates from Section \ref{SecEstimates} do not qualitatively change in the combined geometry and one still requires unfeasible amounts of energy for the potential realization of local inflation bubbles. We remark, however, that while the total energy per equal-time surface remains positive under these specific assumptions, other parameter choices can generally lead to negative energies, in contrast to the positive results for $E_\mathrm{stat}$ in the case of an isolated local inflation bubble, as discussed in Section \ref{GeometricConstruction}.

\section{Discussion}
The central mathematical construction in this work is the local inflation bubble, i.e. a smooth and compactly supported deformation of a given spacetime region. The resulting geometry preserves the large scale structure of the background spacetime, e.g. an asymptotically flat spacetime remains asymptotically flat, and the bubble does not generate gravitational radiation by itself. Although the bubble carries nontrivial (possibly negative) quasi-local gravitational energy, its ADM mass vanishes in the pure bubble case, so that also the global mass of the background is preserved. Our explicit analysis shows that all curvature quantities, particularly the stress-energy tensor components, are bounded and that the bubble introduces no curvature singularities. Interestingly, in the foliation associated to a static observer, the total energy per equal-time surface, which is required to sustain the local expansion, remains nonnegative at all times, even though pointwise energy conditions are generally violated. From an optical perspective, the geometry exhibits a distinctive behaviour. In the central region the light cones steepen during the inflationary phase, strongly suppressing radial propagation and producing short-lived, approximately bifurcating marginal surfaces rather than long-lasting trapping horizons. \\

As a concrete example, we consider the model of a local inflation bubble determined by the $2$-dimensional bump function \eqref{DoubleBump}, providing qualitative estimates of the localization and extent of negative energy densities. Within this specific model, we further provide order-of-magnitude energy estimates across various parameter regimes, indicating that, while the mechanism is self-consistent as a theoretical toy model, the energetic cost is enormous and any practical realization is, at present, out of reach. \\

Finally, we have embedded a Morris-Thorne wormhole into the local inflation bubble and carried out an explicit analysis of the associated curvature quantities and stress-energy contributions. In this combined system, all curvature quantities are finite and all components of the stress-energy components remain bounded. In particular, the latter admit uniform lower bounds by construction. Our results further indicate that, once the exotic-matter-supported wormhole is placed inside the bubble, the total energy per equal-time slice, measured by a static observer, need not remain positive, and the standard pointwise energy conditions continue to be violated. Nonetheless, we identify specific geometric configurations for which the energy density at the throat becomes positive during the local inflation, e.g. when the compactly supported profile is chosen to be sufficiently sharply peaked at the throat. \\

Within any complete theory of quantum gravity that locally gives rise to a quantum foam picture, a local inflation bubble would generically amplify not only wormhole geometries but also the broader spectrum of Planck scale, quantum-induced irregularities, such as microscopic black holes \cite{CMN:2015psbh}, or wave-like curvature fluctuations \cite{JYA:2020ncgw,HJSG:2024ncgw,Vetal:2025ncgw}, see also \cite{PI:2021qgd}, which the bubble would magnify as well. This may be regarded as both a bug and a feature: As the selective production of wormholes is not to be expected, the localization and targeted amplification constitutes an additional challenge, while amplified non-wormhole effects may even pose potential hazards. We remark, however, that the precise effects of local inflation bubbles on quantum-gravitational signatures have not yet been studied rigorously, so that these phenomenological considerations remain purely speculative. \\

Nonetheless, our local construction may provide a useful toy model for amplifying small curvature effects, while preserving analytical control of the background. A particularly favorable aspect is that the construction is asymptotically flat and nonradiative, so that gravitational wave signals observed in the asymptotic region cannot originate from the bubble itself, and must be of different origin. A complete assessment of such scenarios, however, would require a detailed perturbative analysis and computationally demanding numerical studies of signal propagation through the bubble, which are beyond the present scope. \\

The main caveat of our model is, however, that the local inflation bubble still requires exotic stress-energy contributions, at least locally, as pointwise energy conditions are generally violated. While the total energy per constant-time surface, as seen by static observers, is manifestly nonnegative, pointwise violations of the weak and null energy conditions do generally occur and, when quantifying the extent of these violations, the volume integral $\int_{\Sigma_t}(\rho+p_r)\,d\mathrm{vol}_{\Sigma_t}$ may take negative values, driven by a strongly negative radial pressure.  This property highlights the fact that no classical matter model suffices for the realization of local inflation bubbles. In particular, we expect that averaged energy conditions, including the ANEC and the achronal ANEC, are violated in the present setting, although we do not establish this explicitly. We therefore propose that determining whether the isolated local inflation bubble \eqref{BubbleMetric} satisfies or violates the (achronal) ANEC would be of great interest for future work, as it may further restrict the physical viability of our construction. \\

Accordingly, we present the local inflation bubble as a theoretical \emph{Gedankenexperiment}, intended to illustrate a localized geometric mechanism and energetic bookkeeping, rather than to offer precise instructions for engineering exotic spacetimes. Our aim is to examine what would be required, geometrically and energetically, to temporarily amplify Planck scale spacetime structures, while remaining agnostic, if not skeptical, about practical realizability. The model functions as a stress test for earlier enlargement proposals for microscopic wormholes emerging from the quantum foam, most prominently including inflationary expansion \cite{Roman:1993wh} and local false vacuum scenarios \cite{BGG:1987fvb}. The principal features of our refined construction are its compact support in spacetime, and the subsequent preservation of the global asymptotic structure, and isolation of microscopic irregularities from the large-scale geometry. The model also permits a relatively straightforward analysis of inhomogeneous local inflation, facilitating the identification of geometrically special and physically interesting configurations. Altogether, the framework serves as a conceptual laboratory for probing the limits of general relativity by inflating possible geometrical features of the quantum foam picture to macroscopic scales, while explicitly operating in the absence of a complete theory of quantum gravity. \\

Before concluding, we point out a particular direction for future work. For a locally inflated wormhole to be of any practical use, it must withstand interactions with its environment, ranging from test photons to macroscopic probes. This suggests a careful analysis of the (local) stability of a wormhole embedded in a local inflation bubble, as described by the metric \eqref{EmbeddedWormholeMetric}. Prior results within conventional general relativity suggest that static traversable wormholes are often unstable under gravitational perturbations, see \cite{SH:2002whs,GGS:2008whs,KK:2020per}, whereas modified theories of gravity offer new possibilities for obtaining stable traversable wormholes, see, e.g. \cite{SB:2021mod,dFBCdL:2021wh,dFC:2023wh}. In recent developments, further mechanisms, including constructions embedded in the Standard Model or physics beyond the Standard Model, have been proposed to achieve long-lived or stable traversable wormholes, see \cite{MMP:2023wh,MM:2021wh}. It would be interesting to study how such more realistic wormhole solutions behave when embedded in a local inflation bubble. Moreover, the stability and decay properties of false vacuum bubbles in the presence of static traversable wormholes have been studied in asymptotically $\mathrm{AdS}_4$ backgrounds, see \cite{BEFN:2021holo,BFR:2024holo}. It is an important open question to what extent these analyses extend to asymptotically flat spacetimes with compactly supported inflation dynamics, as realized in our construction. More generally, it is particularly intriguing to ask whether spatially and temporally compact external dynamics can contribute to stabilizing the wormhole, at least during the operational window of the local inflation bubble. Our framework provides the basic ingredients for such an analysis, namely, fixed background asymptotics with compactly supported dynamics, explicit control of the stress-energy contributions, and other curvature quantities such as quasi-local mass, null expansions, and a nonnegative total energy per equal-time slice in the foliation associated to static worldlines. Our conjecture is that stability requirements impose additional constraints both on the local inflation bubble and on the wormhole itself, thereby further restricting the set of physically viable configurations. \\

\section*{Acknowledgements}
The authors are very grateful to Maria Alberti for fruitful comments on various details and to an anonymous reviewer for pointing out a mistake in a previous version of the manuscript.

\section*{Funding}
This work was supported by the Open Access Publishing Fund of Leipzig University. The work of PD was funded by the Deutsche Forschungsgemeinschaft (DFG) under Grant No. 406116891 within the Research Training Group (RTG) 2522: "Dynamics and Criticality in Quantum and Gravitational Systems".

\appendix
\section{Exact Results for the Embedded Wormhole Metric}\label{AppendixCurvatureResults}
In the appendix, we briefly collect the explicit results for the curvature quantities associated to the metric
\begin{equation*}
ds^2 = -dt^2 + e^{2f(t,l)} \left( dl^2 + \left(s_0^2 + l^2 \right)\, d\Omega^2 \right),
\end{equation*}

\noindent obtained with the help of computational software \cite{Mathematica,xAct,xCoba,xPerm,xTensor}. Note that the special case $s_0 \rightarrow 0$ recovers the pure local inflation bubble in Minkowski spacetime, without the presence of a wormhole. \\

First of all, we find the nonvanishing Christoffel symbols
\begin{align*}
\Gamma^t_{ll} &= e^{2f} (\partial_t f), & \Gamma^t_{\vartheta\vartheta} &= e^{2f}\left(s_0^2 + l^2 \right) (\partial_t f), \\
\Gamma^t_{\varphi\varphi} &= e^{2f} \left(s_0^2 + l^2 \right) \sin^2\vartheta\, (\partial_t f), & \Gamma^l_{tl} &= \Gamma^\vartheta_{t\vartheta} = \Gamma^\varphi_{t\varphi} = (\partial_t f), \\
\Gamma^l_{ll} &= (\partial_l f), &
\Gamma^l_{\vartheta\vartheta} &= -\left( l + \left(s_0^2 + l^2 \right) (\partial_l f) \right), \\
\Gamma^l_{\varphi\varphi} &= -\left( l + \left(s_0^2 + l^2 \right) (\partial_l f) \right)\sin^2\vartheta, &
\Gamma^\vartheta_{l\vartheta} &= \Gamma^\varphi_{l\varphi} = \frac{l}{s_0^2 + l^2} + (\partial_l f), \\
\Gamma^\vartheta_{\varphi\varphi} &= -\sin\vartheta \, \cos\vartheta, &
\Gamma^\varphi_{\vartheta\varphi} &= \cot\vartheta.
\end{align*}

\noindent Moreover, for the Riemann tensor, we find the components
\begin{align*}
R_{tllt} &= e^{2f} \left( (\partial_t f)^2 + (\partial_t^2 f) \right), \\
R_{t\vartheta\vartheta t} &= e^{2f} \left(s_0^2 + l^2 \right) \left( (\partial_t f)^2 + (\partial_t^2 f) \right), \\
R_{t\vartheta\vartheta l} &= e^{2f} \left(s_0^2 + l^2 \right) (\partial_t \partial_l f), \\
R_{t\varphi\varphi t} &= R_{t\vartheta\vartheta t}\, \sin^2\vartheta, \\
R_{t\varphi\varphi l} &= R_{t\vartheta\vartheta l} \, \sin^2\vartheta, \\
R_{l \vartheta\vartheta l} &= e^{2f} \left( \frac{s_0^2}{s_0^2 + l^2} + l(\partial_l f) + \left(s_0^2 + l^2 \right) \left( (\partial_t^2 f) - e^{2f} (\partial_t f)^2 \right) \right), \\
R_{l \varphi\varphi l} &= R_{l \vartheta \vartheta l} \, \sin^2\vartheta, \\
R_{\vartheta \varphi \varphi \vartheta} &= e^{2f} \left( 2l\left(s_0^2 + l^2 \right) (\partial_l f) + \left(s_0^2 + l^2 \right)^2 (\partial_l f)^2 - e^{2f}\left(s_0^2 + l^2 \right)^2 (\partial_t f)^2 - s_0^2 \right)\sin^2\vartheta,
\end{align*}

\noindent while the remaining nonvanishing components are obtained by symmetry relations. Similarly, we obtain the nonvanishing Weyl tensor components

\begin{align*}
W_{tltl} &= \frac{1}{3} \left( (\partial_l f)^2 - (\partial_l^2 f) + \frac{l}{s_0^2 + l^2}(\partial_l f) - \frac{2 s_0^2}{\left(s_0^2 + l^2 \right)^2} \right), \\
W_{t\vartheta t\vartheta} &= \frac{1}{6} \left( \left(s_0^2 + l^2 \right)(\partial_l^2 f) - \left(s_0^2 + l^2 \right)(\partial_l f)^2 - l(\partial_l f) + \frac{2 s_0^2}{s_0^2 + l^2} \right), \\
W_{t\varphi t\varphi} &= W_{t\vartheta t\vartheta} \,\sin^2\vartheta, \\
W_{l\vartheta l\vartheta} &= \frac{e^{2f}}{6} \left( \left(s_0^2 + l^2 \right) (\partial_l f)^2 - \left(s_0^2 + l^2 \right)(\partial_l^2 f) + l (\partial_l f) - \frac{2 s_0^2}{s_0^2 + l^2} \right), \\
W_{l\varphi l\varphi} &= W_{l\vartheta l\vartheta} \,\sin^2\vartheta, \\
W_{\vartheta\varphi\vartheta\varphi} &= \frac{1}{3} e^{2f} \left( \left(s_0^2 + l^2 \right)^2(\partial_l^2 f) - \left(s_0^2 + l^2 \right)^2(\partial_l f)^2 - l \left(s_0^2 + l^2 \right) (\partial_l f) + 2 s_0^2 \right).
\end{align*}

\noindent Subsequently, the nonvanishing components of the Ricci tensor explicitly take the form
\begin{align*}
R_{tt} &= -3\left( (\partial_t f)^2 + (\partial_t^2 f) \right), \\
R_{tl} &= -2 (\partial_t \partial_l f), \\
R_{ll} &= e^{2f}\left( 3(\partial_t f)^2 + (\partial_t^2 f) \right) - 2 (\partial_l^2 f) - \frac{2l}{s_0^2 + l^2} (\partial_l f) - \frac{2s_0^2}{\left(s_0^2 + l^2 \right)^2}, \\
R_{\vartheta\vartheta} &= e^{2f}\left(s_0^2 + l^2 \right)\left( 3(\partial_t f)^2 + (\partial_t^2 f) \right) - \left(s_0^2 + l^2 \right)(\partial_l^2 f) - \left(s_0^2 + l^2 \right)(\partial_l f)^2 - 3l (\partial_l f), \\
R_{\varphi\varphi} &= R_{\vartheta\vartheta} \; \sin^2 \vartheta,
\end{align*}

\noindent so that the Ricci scalar explicitly computes as
\begin{align*}
R &= 3\left( 2(\partial_t f)^2 + (\partial_t^2 f) \right) \\
&\hspace{0.5cm} - e^{-2f}\left( 2(\partial_l^2 f) + (\partial_l f)^2 + \frac{4l}{s_0^2 + l^2} (\partial_l f) + \frac{2s_0^2}{\left(s_0^2 + l^2 \right)^2}\right).
\end{align*}

\noindent Consequently, these curvature quantities are used to compute the Einstein tensor, for which we obtain the explicit components
\begin{align*}
G_{tt} &= 3(\partial_t f)^2 - e^{-2f} \left( \frac{s_0^2}{\left( s_0^2 + l^2 \right)^2} + \frac{4l}{s_0^2 + l^2}(\partial_l f) + (\partial_l f)^2 + 2\,(\partial_l^2 f) \right). \\
G_{ll} &= \frac{2l}{s_0^2 + l^2} (\partial_l f) + (\partial_l f)^2 - e^{2f}\left( 3(\partial_t f)^2 + 2(\partial_t^2 f) \right) - \frac{s_0^2}{\left(s_0^2 + l^2 \right)^2},\\
G_{\vartheta\vartheta} &= \frac{s_0^2}{s_0^2 + l^2} + l(\partial_l f) + \left(s_0^2 + l^2 \right) (\partial_l^2 f) - e^{2f}\left(s_0^2 + l^2 \right)\left( 3(\partial_t f)^2 + 2(\partial_t^2 f) \right),\\
G_{\varphi\varphi} &= G_{\vartheta\vartheta} \; \sin^2\vartheta,
\end{align*}

\noindent together with the $tt$-component given by \eqref{T00EmbeddedWormhole}. Lastly, the NP null tetrad associated to the metric \eqref{EmbeddedWormholeMetric} is explicitly given by

\begin{align*}
    l^a &= \frac{1}{\sqrt{2}}\left(\frac{\partial}{\partial t} \right)^a + \frac{e^{-f}}{\sqrt{2}}\left(\frac{\partial}{\partial r} \right)^a, \\
    n^a &= \frac{1}{\sqrt{2}}\left(\frac{\partial}{\partial t} \right)^a - \frac{e^{-f}}{\sqrt{2}}\left(\frac{\partial}{\partial r} \right)^a,\\
    m^a &= \frac{e^{-f}}{\sqrt{2\left(s_0^2 + l^2 \right)}}\left(\frac{\partial}{\partial\vartheta} \right)^a + \frac{ie^{-f}}{\sqrt{2\left(s_0^2 + l^2 \right)}\, \sin\vartheta}\left(\frac{\partial}{\partial \varphi} \right)^a, \\
    \overline{m}^a &= \frac{e^{-f}}{\sqrt{2\left(s_0^2 + l^2 \right)}}\left(\frac{\partial}{\partial\vartheta} \right)^a - \frac{ie^{-f}}{\sqrt{2\left(s_0^2 + l^2 \right)}\,\sin\vartheta}\left(\frac{\partial}{\partial \varphi} \right)^a,
\end{align*}

\noindent which, by comparison with expressions \eqref{Outgoing} and \eqref{Ingoing}, demonstrates that the propagation of ingoing and outgoing light rays is identical to that in a pure local inflation bubble, corresponding to the limit $s_0 \to 0$. \\

{\footnotesize
\bibliographystyle{unsrt}
\bibliography{localinflationbubbles}}

\end{document}